\documentclass[10pt,conference,review]{IEEEtran}
\IEEEoverridecommandlockouts

\usepackage{amsmath}
\usepackage[T1]{fontenc}

\usepackage{enumerate}
\usepackage{fancyvrb}
\usepackage{amssymb}
\usepackage{color}
\usepackage{amsfonts}
\usepackage{amssymb}
\usepackage{amsthm}

\usepackage{listings}
\usepackage{graphicx}
\usepackage{algorithm}
\usepackage{lipsum}
\usepackage{pifont}
\usepackage{verbatim}
\usepackage{multirow}
\usepackage{framed}
\usepackage{caption}
\usepackage{subcaption}
\usepackage{soul}
\usepackage{mdframed}
\usepackage{tikz}
\usepackage{tikzscale}
\usetikzlibrary{arrows.meta}
\usetikzlibrary{arrows}
\usetikzlibrary{positioning}
\usetikzlibrary{shadows}
\usepackage{balance}
\usepackage{paralist}
\usepackage{booktabs}
\usepackage{rotating}
\usepackage[font={small,bf}]{caption}
\usepackage{multicol}
\usepackage{algpseudocode}
\usepackage{fancyvrb}
\usepackage{soul}
\usepackage{xcolor}
\usepackage{newverbs}
\usepackage{hyperref}

\newcommand{\commentxh}[1]{{\color{blue} \sf (XH: #1)}}
\newcommand{\ignore}[1]{}
\newcommand{\name}{ConfRL}

\newcommand{\revision}[1]{{#1}}

\begin{document}
\setlength{\textfloatsep}{0.1cm}
\setlength{\floatsep}{0.1cm}

\title{Automated Performance Tuning for Highly-Configurable Software Systems}

\author{\IEEEauthorblockN{Xue Han}
\IEEEauthorblockA{\textit{Department of Computer Science} \\
\textit{University of Southern Indiana}\\
Evansville, USA \\
xhan@usi.edu}
\and
\IEEEauthorblockN{Tingting Yu}
\IEEEauthorblockA{\textit{Department of Computer Science} \\
\textit{University of Kentucky}\\
Lexington, USA \\
tyu@cs.uky.edu}
}

\maketitle

\begin{abstract}
Performance is an important non-functional aspect of the software requirement. 
Modern software systems are highly-configurable and 
misconfigurations may easily cause performance issues. 
A software system that suffers performance issues may exhibit low
program throughput and long response time.
However, the sheer size of the configuration space makes it challenging for 
administrators to manually select and adjust the configuration options to 
achieve better performance. 
In this paper, we propose \name{}, an approach
to tune software performance automatically.
The key idea of \name{} is to use reinforcement learning 
to explore the configuration space by a trial-and-error approach 
and to use the feedback received from the environment to 
tune configuration option values to achieve better performance.  
To reduce the cost of reinforcement learning,
\name{} employs sampling, clustering, and dynamic state reduction
techniques to keep states in a large configuration space manageable. 
Our evaluation of four real-world highly-configurable server programs shows that
\name{} can efficiently and effectively guide software systems to achieve a higher 
long-term performance.
\end{abstract}


\section{Introduction}
\label{sec:intro}

Modern computer systems are highly-configurable, allowing users 
to customize a large number of configuration options
\footnote{A.k.a., configuration parameters. 
Configuration used in this context is not to be confused with software configuration management (SCM).}
to meet their specific goals.  The complexity of the configuration space and the 
sophisticated constraints among configurations
may easily lead to performance issues. 
Recent studies have shown that performance problems caused by misconfiguration are 
still prevalent~\cite{Attariyan12,Han16,Jin12}.  
Performance issues can cause significant performance degradation which leads to 
long response time and a low program throughput~\cite{bugzilla,Jin12,NistorMSR}.


When a performance problem occurs 
(e.g., a significant slowdown with HTTP responses in a web server), 
system administrators and developers may need to 
re-configure the system to find a configuration 
setting for better performance.
However, it is not an easy task to figure out the best settings for 
a system with a large number of configuration options. 
\revision{For example, the latest version of Apache HTTP Server (Apache) has
600+ configuration options (directives). 
Even for domain experts, it is not an easy task to 
configure the software system to get the best
performance.} For example, as one experienced
user complained in HBase Bug \#13919\footnote{https://issues.apache.org/jira/browse/HBASE-13919},
``There are current many settings that influence how/when an HBase client times out.
This is hard to configure, hard to understand, and badly documented.''
Besides, manually changing the configuration can be tedious, inefficient, and 
impractical. 
For instance, in the case of a web server, 
the volume of the web page request level changes at different times
of the day. It is not practical to ask administrators to change configurations 
to keep up with the level of web request changes.


\revision{The goal of this research is to develop an approach, \name{},  
that can automatically select and tune configuration options in 
response to the environment dynamics~\cite{chung2006case} to achieve 
better performance.
 \name{} is intended to be used by system administrators
and developers 
to tune performance through configurations.} 
The key idea of \name{} is to use reinforcement
learning (RL) techniques to automate performance configuration tuning. 
RL is a process of interacting with a dynamic environment
to generate the optimal control policy on what actions to take for a given state.
\revision{Therefore, we can formulate the task
of performance tuning as a RL problem
in which the optimal policy refers to generate a configuration
for higher performance.}
%
The main benefit of RL is that it does not require domain knowledge of
the software system and can update optimal policies
continuously in the long run.  
%

\name{} consists of two stages: performance-influential configuration option ranking
and Q-Learning~\cite{wikiQlearning}. The first stage identifies configuration options that potentially
influence the system's performance. 
The enormous configuration space leads to a huge number of states
that RL must explore, thus apply RL directly to a system with a large number of
configuration options hardly scales. 
To address this challenge, \name{} uses a clustering method to identify
performance-influential options. 
 
The second stage of \name{} uses Q-Learning for finding a policy to
guide on what actions to take in a given state to 
achieve higher performance gains. 
There are two challenges in this stage. 
First, even with fewer options to choose from, 
the configuration space may still be enormous~\cite{valmari1996state} at runtime. 
To address this challenge, we utilize 
adaptive value generation and dynamic state merging techniques
to reduce the runtime reinforcement learning states.
Another challenge is the inconsistent readings of the 
performance measurements (e.g., program throughput such as web page requests per second) 
across multiple executions with the same inputs and configurations. 
\revision{Such inconsistencies 
may disturb the results of the calculated rewards in Q-Learning. 
We address this challenge by caching performance measurements 
for visited states to obtain consistent readings.}

\name{} differs from existing work~\cite{jin2014preffinder,Swanson14} on configuration tuning. 
For example, PrefFinder~\cite{jin2014preffinder} uses information retrieval (IR) from 
static documentation to automatically  find user preferences for correcting the 
configuration of a running system. In contrast, \name{} focuses on addressing
performance problems and does not assume the availability of any documentation,
which may be incomplete or out of date.   
\revision{In comparison to REFRACT~\cite{Swanson14},
a self-adaptive approach to avoid software failures, first, our goal is to 
achieve higher long-term performance gains, which is intrinsically different than software failures avoidance.

Second, sampling alone does not output a policy to guide configurations.
Instead of relying on sampling techniques to find workarounds, we use reinforcement 
learning to take advantage of past interactions between the agent and the environment 
to guide the subject systems toward long-term software performance gains.}

We evaluate \name{} on four popular real-world C/C++ programs. 
Our results show that \name{} is effective in improving performance
through automated configuration tuning within a reasonable time. 
Compared to a random tuning approach, \name{} is 
up to 30\% more effective in achieving performance gains.
Moreover, the optimization techniques employed by \name{} 
significantly reduce the number of states for reinforcement learning up to 82.5\%
and thus require less time (20.5 hours) to converge.
%

In summary, our paper makes the following contributions:

\begin{itemize}
 \item \revision{An automated tool that selects and tunes configuration options to achieve
 long-term performance gains.}
 
 \item \revision{A scalable approach that improves reinforcement learning with 
 state reduction techniques to accelerate performance tuning.} 
 
 \item  Practical implementation and empirical evidence to show that the approach can 
    effectively and efficiently improve performance in real-world server programs.

\end{itemize}

This paper is structured as follows.
In Section~\ref{sec:background}, we introduce the technical background.
In Section~\ref{sec:FormulateRL}, we discussion the formulation of \name{} as a RL problem.
We then present the design of \name{} in Section~\ref{sec:approach}.  
Our empirical study and results are presented in Sections~\ref{sec:study} and \ref{sec:results}, 
followed by a discussion of threat to validity in Section~\ref{sec:discussion}.  
We present the related work in Section \ref{sec:related}, and then conclude the paper in
Section~\ref{sec:conclusion}.
\section{Background}
\label{sec:background}

In this section, we  introduce the background of reinforcement learning (RL) 
and  discuss how to model automated performance tuning 
as a RL task.

\subsection{Tuning Performance Configuration}

\begin{figure}[t]
 \includegraphics[scale=0.35]{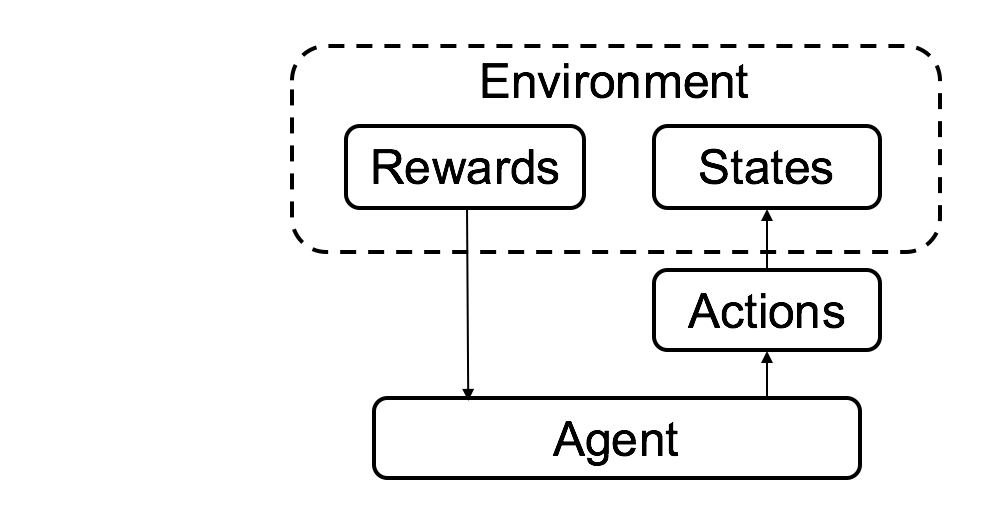}
  \caption{Reinforcement Learning in A Nutshell}
  \label{fig:RL}
\end{figure}

Reinforcement learning (RL)~\cite{rlearning}, 
as illustrated in Figure~\ref{fig:RL}, is the procedure of learning from interactions 
between an agent and the environment to determine the best action to take under any given state to 
achieve the maximum long-term rewards~\cite{sutton1998introduction}.
\revision{After the agent initiates an action, the environment reacts to the action by
transiting the agent to a different state.}
\revision{Depending on the current and previous states, 
the environment grants rewards to the agent.
This cycle goes on iteratively until the learning procedure terminates.
The output of RL is a policy that guides the agent to take an action that maximizes long-term rewards
based on its current state.
The value of an action associated in a state
is computed by a function that estimates the sum of 
future rewards by taking this action.
The agent performs 
trial-and-error interactions with the environment to obtain 
rewards. Therefore, the optimal policy is to select the 
action that maximizes the reward in each state.}

\revision{The task of performance tuning fits 
into the RL framework naturally. 
Each configuration (i.e., a combination of configuration option values)
represents a RL state.
When performance tuning occurs, (i.e., issue an action to change configuration), 
an action receives a reward based on performance measurements. 
Given sufficient interactions with the environment, RL obtains 
an estimation of how good an action is for the current 
configuration (i.e., state). 
}

\subsection{Reinforcement Learning Techniques}
Reinforcement learning comes in a few different forms
depending on what is available to the problem
(e.g., full knowledge of the environment such as the transition function). 
\revision{We discuss briefly two of the most widely adopted methods and 
explain why we select the Q-Learning method for our problem.}

{\bf Markov Decision Process.}
The basic form of a reinforcement learning problem is encapsulated 
as the Markov Decision Process (MDP)~\cite{puterman2014markov}.
Formally, MDP is used to describe an environment for reinforcement learning 
where the environment is fully observable. 
MDP consists of a finite set of states, a finite set of actions,
a state transition matrix: $P_{ss'}= \mathbb{P}[S_{t+1} = s' | S_t = s]$~\cite{dsilver15},
a reward function $\mathcal{R}$, and a discount factor $\gamma$.
A state has Markov property if and only if each state captures the information from
all past states that lead to the current state.
A policy $\pi$ gives the probability to take an action given a state $s$:
$\pi(\alpha|s) = \mathbb{P}[A_t = \alpha | S_t = s]$.
\revision{The action-value function $q_\pi(s,\alpha)$ is the expected total rewards for taking 
an action $\alpha$ following the policy $\pi$ in state $s$.}
The goal of solving the MDP problem is to find the optimal action-value function:
$q_*(s,\alpha) = \max\limits_{\pi} q_\pi(s,\alpha)$.

%
{\bf Model-Free RL.}
If a problem can be modeled as MDP, it can be solved analytically 
through value-iteration and policy-iteration algorithms.
Most real-world problems cannot be formulated as MDP since the environment is not fully observable. 
It is also difficult to describe the rules in a dynamic environment, so the MDP transition function is unknown.
There are a set of techniques to estimate the action-value function of an unknown MDP, such methods are
referred to as Model-Free reinforcement learning algorithms~\cite{dsilver15}.
The Q-Learning method~\cite{wikiQlearning} is one type of Model-Free learning algorithm. 
It seeks to learn a policy to maximize the total award.
\name{} uses Q-Learning as its reinforcement learning algorithm.
Q-Learning is based on the Bellman Optimality Equation~\cite{bradtke1995reinforcement}:
\begin{center}
$q_*(s,\alpha) = \mathcal{R}_s^\alpha + \gamma \sum\limits_{s' \in S} \max\limits_{\alpha'} q_*(s',\alpha')$.
\end{center}
\revision{The formula contains two parts: $\mathcal{R}_s^\alpha$ is the immediate reward, 
$\sum\limits_{s' \in S} \max\limits_{\alpha'} q_*(s',\alpha')$ is 
the expected future reward to take an action $\alpha$ in state s,
and $\gamma$ is the discount factor that determines how much RL should value the future
reward.}

%
{\bf $\epsilon$-Greedy Exploration.}

\revision{As RL explores the environment, it takes advantage of 
its experience interacting with the environment.
Internally, RL maintains a state-action table to keep track of rewards
received by taking specific actions in a state. This helps 
RL to select actions to achieve higher performance gains. 
However, in the early stage of exploration, sticking to the state-action table completely 
may restrict the number of states that RL could have visited. 
There may be a chance that RL could have achieved 
higher performance by visiting different states. 
As such, RL introduces some randomness into the process.
By convention,
the degree to which RL acts randomly is denoted as $\epsilon$
with a range between 0 (i.e., no random actions at all) and 1.}

\section{Problem Formulation}
\label{sec:FormulateRL}

\begin{figure*}[t!]
\center
\includegraphics[scale=0.5]{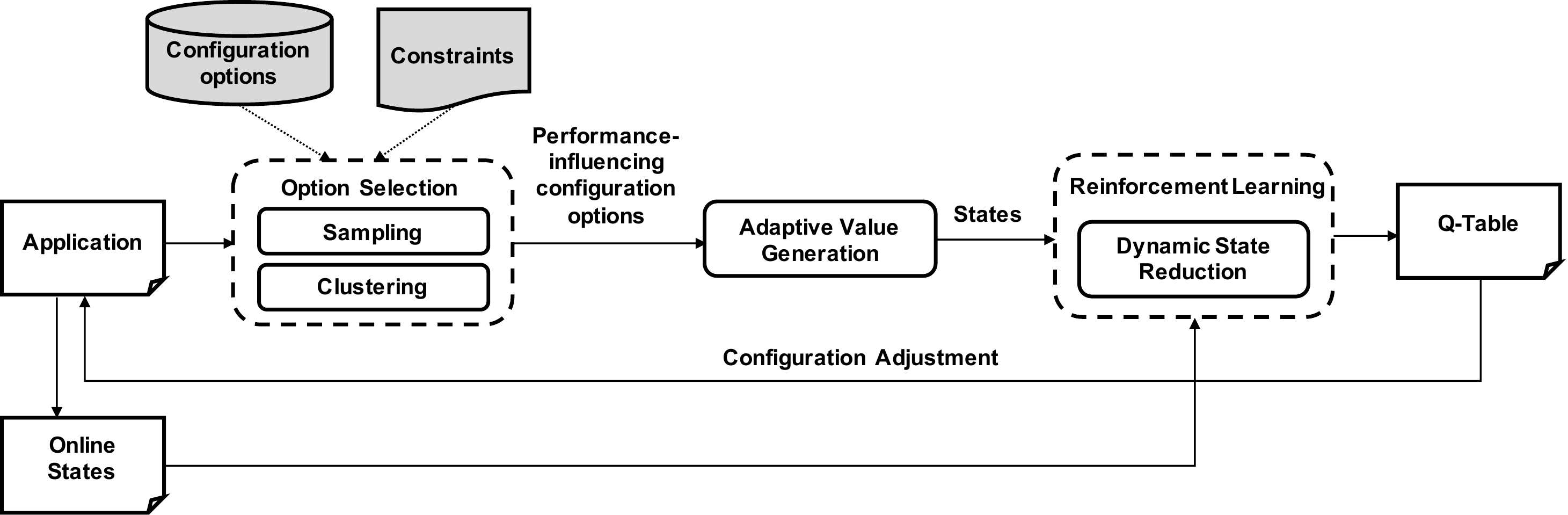}
  \caption{\name{} Overview}
  \label{fig:overview}
\end{figure*}
 
We demonstrate how to formulate and solve the problem of performance tuning 
with RL using an example of the Apache web server. 
\revision{Table~\ref{tab:ops} shows the configuration options used in the example.
Column ``Name'' lists the names of the selected configuration options. 
Column ``Type'' lists the option value types: binary  (B) or numerical (N).
Column ``Range'' lists the configuration option value ranges.
Column ``Constraints'' lists imposed constraints on configuration options.
The configuration option value ranges and constraints are manually extracted from 
the documentation of the subject programs.}

\begin{table}[t]
\centering
\scriptsize
\setlength{\tabcolsep}{4pt}
\caption{\label{tab:ops} Configuration Options}
\begin{tabular}{|c|c|c|c|} \hline
{\em Name} & {\em Type} & {\em Range} & {\em Constraints} \\
\hline 
KeepAlive & B & OFF | ON & {Constraints Not Applicable}\\
\hline
MaxClients & N & [1,512] & MaxClients < ServerLimit\\
\hline
StartServers & N & [1,100] & StartServers < MaxSpareThreads \\
\hline
{ThreadsPerChild} & {N} & {[1,128]}
& ThreadsPerChild * StartServers < MaxClients\\
\hline
\end{tabular}
\begin{center}
\end{center}
\normalsize

\end{table}

{\bf State.}

A state is encoded as an instance of the subject program's configurations.
For example, Table~\ref{tab:apaStates} illustrates five states in Apache.
\revision{Each state is a combination of the current configuration option values (Columns ``Option'').}
The ``State'' column lists the state ID.
The default values of configuration options are used as the 
initial state (S1). We discuss the impact of choosing different initial states
in Section~\ref{sec:discussion}.
The ``Meas.'' column lists the performance measurements. 

 It is a metric to quantify program performance. 
For web servers, we measure the number of concurrent web page requests 
 per second~\cite{ab}; for database servers, we measure the number of transactions per second~\cite{dbt2}.
The ``Action'' column lists the next actions to be performed on the subject program.
The ``Rewards'' column lists the immediate performance reward. 
The reward is used to populate the state-action table, 
a.k.a, the Q-Table.
The ``State'' column in Table~\ref{tab:qTable} lists the visited RL states.
\revision{Column ``A1'' to ``A8'' list eight actions associated with four options 
used in Table~\ref{tab:apaStates}.
Each cell in Table~\ref{tab:qTable} lists the immediate performance reward
after taking an action in the corresponding state. By default, the immediate performance
reward is set to 0.}

\begin{table}[t]
\centering
\footnotesize
\setlength{\tabcolsep}{4pt}
\caption{\label{tab:apaStates} Apache RL States}
\begin{tabular}{|c|c|c|c|c|c|c|c|} \hline
\multirow{2}{*}{\em State} & 
\multicolumn{4}{c|}{\em Options} & \multirow{2}{*}{\em Meas.} & 
\multirow{2}{*}{\em Action} & \multirow{2}{*}{\em Rewards} \\
& K.A. & M.C. & S.S. & T.P.C. & & &\\
\hline
S1 & OFF & 102 & 12 & 3 & 10 r/s & A1& 1 \\
\hline 
S2 & \textbf{ON} & 102 & 12 & 3 & 20 r/s & A5 & 0.25 \\
\hline
S3 & ON & 102 & \textbf{32} & 3 & 25 r/s & A7 & 0.2 \\
\hline
S4 & ON & 256 & 32 & \textbf{4} & 30 r/s & A6 & -0.17\\
\hline
S5 & ON & 256 & \textbf{16} & 4 & 20 r/s & A5 & N/A\\
\hline
\end{tabular}

\begin{center}
M.C.: MaxClients; S.S.: StartServers; T.P.C: ThreadsPerChild; \\
Meas.: Measurement; r/s: requests per second; \\
\end{center}
\normalsize
\end{table}


%
\begin{table}[t]
\centering
\footnotesize
\caption{\label{tab:qTable} State-Action Table}
\begin{tabular}{|c|c|c|c|c|c|c|c|c|} 
\hline
{\em State} & {\em A1} & {\em A2} 
 & {\em A3} & {\em A4} & {\em A5} 
 & {\em A6} & {\em A7} & {\em A8} \\
\hline
S1 & \textbf{1} & 0 & 0 & 0 
& 0 & 0 & 0 & 0 \\
\hline
S2 & 0 & 0 & 0 & 0 
& \textbf{0.25} & 0 & 0 & 0 \\
\hline
S3 & 0 & 0 & 0 & 0 
& 0 & 0 & \textbf{0.2} & 0 \\
\hline
S4 & 0 & 0 & 0 & 0 
& 0 & \textbf{-0.17}& 0 & 0 \\
\hline

\end{tabular}
\begin{center}
\end{center}
\normalsize
\end{table}

{\bf Action.}
An action is an update issued by \name{} to modify an individual configuration 
option value. For numerical option types, an action can be 1) increasing an option value;
2) decreasing an option value. For binary options types,  an action can be 
1) setting a binary option value to $True$\footnote{Depending on the specific subject, the actual value could also be ``ON'', ``1'', ``True''.}; 
2) setting a binary option value to $False$.
\revision{Action is indexed and encoded by an integer number. Each integer is mapped to a specific operation 
to the option, as shown in the ``Action'' column of Table~\ref{tab:apaStates}. 
In this example,  action one (A1) is mapped to set the first configuration option 
\texttt{KeepAlive} to ``ON''.  
Action five (A5) is mapped to increase the value of the 
third configuration option \texttt{StartServers}.}

{\bf Reward.}
\label{subs:reward}
A reward is calculated based on performance measurement,
\revision{for example, web servers use the number of concurrent 
web page requests per second.
\name{} first obtains the performance measurement 
under the current configuration options as M$_C$.
The subject program follows an action to enter the next state with
performance measurement M$_N$.
The reward is the relative difference between M$_N$ and M$_C$:
(M$_N$ - M$_C$) / M$_C$, 
the normalization puts a large measurement range on the same scale.

Since obtaining the performance measurement 
for each execution is time-consuming and that
the measurements may be inconsistent due to environment
dynamics, \name{} caches the performance measurement 
for each state.}
The cache can 1) speed up the process of 
getting performance measurement, and 2) guarantee the performance measurement is consistent throughout the 
learning process. The details of the performance 
caching technique will be discussed in Section~\ref{sec:measurePerf}.

\section{Design of \name{}}
\label{sec:approach}

Figure~\ref{fig:overview} gives an overview of \name{}. 
\revision{\name{} takes as inputs a program, its
configuration options, and the associated constraints.  
In stage I, \name{} selects options
that influence the program's performance to reduce the RL states
and thus the learning cost. \name{} employs sampling and clustering
techniques to rank the performance-influential options and assign 
them with appropriate weights.
\name{} differs from prior work~\cite{jamshidi2018learning,nair2018finding} 
on performance-influential configuration option identification
in that it is self-contained and it does require any prior knowledge about the software system. 
In stage II, \name{} uses an \emph{adaptive value generation} method to systematically
generate configuration option values to cover a wide value range.
\name{} calculates the reward based on the 
performance measurement obtained in each state and uses the reward
to build the Q-Table.}

\revision{To reduce the learning cost, \name{}  employs a \emph{state merging} technique
to combine reinforcement learning runtime states that
share the same performance measurements.

This can effectively reduce the size of the state space and
lead to faster learning.
In the end, \name{} outputs a Q-Table to reflect the latest 
interactions between the agent and the environment.

This procedure goes on until when reaching the 
stopping criterion. For instance, the
learning procedure stops after 24 hours in our experiment.}

\subsection{The \name{} Algorithm}
Algorithm~\ref{alg:rl} illustrates the pseudocode of \name{}.
The algorithm takes as input the subject's configuration options with default 
values and outputs a Q-Table. 
The algorithm starts with a $t$--way sampling technique~\cite{yilmaz2006covering}
to get the clustering training data (Line 1).
Then, the ranking of performance-influential options is generated from clusters (Line 2).
\name{} creates an action-value table to store the calculated rewards (Line 3). 
Next, the algorithm initiates the Q-Learning algorithm (Line 4). 
There are two hyper-parameters involved in the Q-Learning algorithm: 
the learning rate $\alpha$ and the discount factor $\gamma$. 
The learning rate ``determines to what extent newly acquired
information overrides old information''~\cite{wikiQlearning},
which controls how fast the reinforcement learning converges.
\revision{The discount factor weighs on how RL perceives future rewards.
$\gamma$ has a value between 0 and 1, where 0 indicates RL takes 
only the immediate rewards without considering for a long-term reward (i.e., the 
expected reward for taking an action onwards)
and 1 indicates RL favors the learning towards 
long-term rewards.}

\revision{The $\epsilon$-Greedy explorer (Line 5) is used to control the degree to which RL follows the original policy. 

It is a discrete explorer that follows the greedy policy while maintaining 
a chance to take a random action to explore unknown states.
The randomness is controlled by $\epsilon$. 
The algorithm instantiates the environment and the agent objects (Lines 7--8,)
and then starts the reinforcement iterations inside a while loop (Line 9).
The subject program is reset to the original state at the beginning of each iteration
and exploits what it has learned from past in the next iteration.

When the iteration starts, the \texttt{DoInteractions} function (Line 10) controls interactions between
the agent and the environment. 
This is also where the adaptive value generation starts.
The algorithm dynamically reduces the value of $\epsilon$ (Lines 11--13) to converge faster.} 
State merging (Line 15) is used at the end of the learning iteration to reduce
the number of runtime states. 
The learning process terminates when it reaches a time-threshold.

\begin{algorithm}[t]
\small
\caption{\textbf{\name{}}}
\label{alg:rl}
\begin{algorithmic}[1]
\raggedright
\Require Initial set of configuration options ($O$)
\Ensure Q-Table
 \State $trainingData = tWaySampling(O)$
 \State $rankedOps = PerfClustering(trainingData)$
 \State $avt = ActionValueTable()$
 \State $learner = QLearner(\alpha,\gamma)$
 \State $explorer = EpsilonGreedyExplorer(\epsilon)$
 \State $learner.SetExplorer(explorer)$
  \State $env = ConfigLearnEnv(rankedOps)$ 
 \State $agent = LearningAgent(avt, learner,env)$

\While{$TRUE}$
 \State ${\tt DoInteractions ()}$       
 \If{$TriggerEpsilonUpdate() == TRUE$}
   \State $\epsilon = UpdateEpsilon()$
   \State $explorer = UpdateExplorer()$
 \EndIf
 \State $env.StateMerging()$
\EndWhile
\State
\Function{StateMerging}{} 
 \State $perfM = MeasurePerf()$
 \If{$perfM \in PerfState.keys()$}
  \State $masterState = PerfState[perfM]$
  \State $UpdatePerfState(masterState,stateID)$
 \Else
  \State $PerfState.add(perfM,stateId)$ 
 \EndIf
\EndFunction

\end{algorithmic}
\end{algorithm}

\subsection{Ranking Performance-Influential Configurations}
\label{sec:cClustering}

This pre-processing step is used to identify the performance-influential
 configuration options from the target configuration option space.
A $t$--way covering array samples the set of configurations in such a way that all possible 
$t$--way combinations of options appear at least once.
Based on a previous study~\cite{czerwonka2006pairwise}, a 3-way covering array is adequate to cover 90\% of
interactions between options for configuration sampling.
Configuration options are grouped based on their functional units (a.k.a modules), 
for instance, configuration options under the core and mpm\_common modules 
are grouped in Apache for testing out web page requests.

\revision{A 3-way covering array is used to conduct configuration sampling in \name{}.
The sampled configuration options are executed
and the performance measurements are recorded}. 
The intuition is that performance-influential options tend to influence performance through
drastic value changes. The configuration option often goes beyond a threshold value
to change program performance significantly.
For instance, in Apache Bug \#54852\footnote{https://bz.apache.org/bugzilla/show\_bug.cgi?id=54852}, 
only when the \texttt{StartServers} option is set to a relatively 
large number (i.e., \texttt{StartServers} = 64) will it cause a slowdown in Apache.
Clustering method naturally distinguishes such options by putting data instances in 
the corresponding clusters.

The popular K-Means~\cite{kmeans} clustering method is used for its ease of interpretation and implementation.  
The clusters with the highest and lowest mean performance measurement are selected.
Because the performance measurement is a direct result of the configuration options,
options in each cluster may be used to describe the characteristics of the 
underlying clusters. 
Therefore, the mean value of each configuration option is used
to calculate the difference between datasets.
We measure the difference in option value changes.
And the configuration options are ranked in a descending order based on the difference.
\revision{Because only a small subset of configuration options can cause performance degradations~\cite{Han16},
\name{} prioritizes the top-10 options by assigning a bigger weight 
so that they have a higher chance to be selected in the reinforcement learning.}


%
\subsection{Generating Configuration Option Values}
\label{sec:avGeneration}
Extensively selecting all the legal values within the configuration option value range 
is not practical. In this step, we want to select option values that cover a decent range
without exhaustively trying out every possible value. 
The value range for each configuration option is extracted from 
the subject documentation in the form of [OPT$_{MIN}$, OPT$_{MAX}$]. 
OPT$_{MAX}$ is set to twice the size of the recommended
max value to ensure a wide value range coverage to expose 
performance problems~\cite{xiao2013context}.

\name{} utilizes an adaptive value generation strategy to generate option values.  
Unlike the sampling method where data points are calculated before any execution,
RL is a dynamic process which requires the option values to be generated on the fly 
with respect to the action.
Initially, each option starts with the default value. 
\revision{Because the impact of configuration option values on the program performance is not continuous, 
performance bugs are often triggered by going above or below a threshold value~\cite{Jovic11}. 
Therefore, it is not necessary to generate option values in a continuous manner. 
Similar to the binary search algorithm, 
the value adjustment is proportional to the power of two.}
Therefore, the maximum number of option
value choices is bounded by log(OPT$_{MAX}$). 
For instance, the MaxClients option has a value range of [1, 1024].
As such, we have at most eleven values choices for MaxClients:
$2^0, 2^1, ..., 2^9, 2^{10}$.
\name{} makes sure the selected value satisfies all
constraints imposed on the option using the python-constraint library~\cite{niemeyer2017python}.

\subsection{Reducing Runtime States}
\label{sec:dsReduction}

\revision{The adaptive value generation strategy can significantly reduce 
the number of individual option values without losing adequate
coverage for exposing performance problems.}
However, as the number of configuration options used in the learning process increases, 
it still poses a challenge to handle a large number of option value combinations, aka, 
the RL states.
To further reduce the number of RL states, 
\name{} uses a dynamic state reduction strategy. 

\revision{At runtime, different states do not always 
lead to different performances.
Many states lead to unnecessary costs in measuring performance
without providing any new insights for the reinforcement learning process.
Only those performance-influential configuration options
tend to result in different performance measurements.
\name{} merges reinforcement learning states that 
share the same runtime performance.}
\name{} implements a cache to store performance measurements for states.
At the end of each reinforcement learning iteration (Line 15 in Algorithm~\ref{alg:rl}), 
a reference list is constructed for state IDs that have 
identical performance measurements (Lines 21--22 in Algorithm~\ref{alg:rl}).
The first such state in each reference list is referred to as the master state, 
the rest of the states in the reference list is referred to as slave states.
In the following reinforcement learning iteration, 
\name{} returns the master state if 
the current state is in the slave state list.

\subsection{Measuring Performance}
\label{sec:measurePerf}
Performance measurements (e.g., execution times, throughputs)
are used to evaluate if one state is better than another state in terms of 
achieving higher performance.
Performance measurement is essential to calculate rewards
in \name{} for Q-Learning (Section~\ref{sec:FormulateRL}).
Within each reinforcement learning iteration, \name{} measures
the performance by executing benchmark tools (e.g., Apache Benchmark, DBT-2).
To provide a reliable and consistent performance measurement, 
the performance measurement of a state is stored upon the first time \name{} 
visits the state. 
\revision{Specifically, the state and its performance measurement 
are stored in a cache.
The cache uses the state ID as the key and the corresponding performance 
as its value. 
In each subsequent reinforcement learning iteration, the performance
of the same state is queried and retrieved directly from the
cache instead of re-running the benchmark utility.
This strategy guarantees the performance of the same state is 
consistent throughout the learning process.
Inconsistent performance measurement has a negative impact 
on the RL reward calculation, therefore confusing \name{} on 
which actions to take in a given state.
The performance measurement cache also reduces 
the overall learning time as benchmark tools can 
take a significant amount of time to collect the performance measurement.}

\subsection{An Example}
\label{subsec:RLEG}

To demonstrate the reinforcement learning design of \name{}, 
we use Table~\ref{tab:apaStates} to show how \name{} works.
\revision{As the learning iteration advances, the Q-Table gets populated and updated
to allow the best action to be returned based on the current state.
Table~\ref{tab:qTable} illustrates the status of Q-Table as the learning progresses.
In this example, we assume the \texttt{StartServers} option has a larger weight than other options. 
Therefore, this option has a higher chance to be selected
by the \name{} (e.g., the \texttt{StartServers} option has been selected in two out of five cases).}

The process starts at the state S1: \{OFF,102,12,3\}. 
S1 has a performance measurement of 10 requests/second (r/s).
In the first interaction, the agent receives the action A1, that is to modify the 
value of the \texttt{KeepAlive} (K.A.) option.
\name{} looks up the configuration option type and confirms
that \texttt{KeepAlive} is a binary option type.
\name{} assigns the value \texttt{ON} to \texttt{KeepAlive}, as such,
Apache is now in a new state S2: \{ON,102,12,3\}.
\revision{After running the benchmark tool, S2 gets a performance measurement of 20 r/s.}

In the meanwhile, \name{} calculates the immediate reward for taking action A1 
in S1 is 1 (i.e., (20-10)/10).
Table~\ref{tab:qTable} gets updated to keep track of the rewards assigned in the state S1.

In the second interaction, the agent receives the action A5 to modify 
the \texttt{StartServers} (S.S.) option. 
\name{} recognizes \texttt{StartServers} as a numerical option type and
calculates the next option value for \texttt{StartServers}.
Besides taking the option's type and its current value (i.e., S.S. = 12), 
\name{} checks the option constraint to make sure that 
all constraints associated with this option 
are still intact (e.g., \texttt{ThreadsPerChild * StartServers < MaxClients} must
hold true for \texttt{StartServers}).
The adaptive value generation adjusts the option value by finding
the first value in the series (i.e., $2^n$) that is larger than 24. 
Therefore, the \texttt{StartServers} option gets a value of 32. 
Apache is now in the state S3: \{ON,102,32,3\}.
The performance measurement of S3 is 25 r/s.
Therefore, the immediate reward for taking action A5 in S2 is 0.25.

In the third interaction, the agent receives the action A7 which is to increase the 
value of \texttt{ThreadsPerChild} (T.P.C.).
\texttt{ThreadsPerChild} gets a value of 4. When the \name{} validates the 
constraints, it no longer holds: \texttt{ThreadsPerChild (4) * StartServers (32) > MaxClients (102)}.
\name{} uses the Constraint Satisfaction Problems (CSPs) solver.
The constraint is passed to the solver as a lambda function 
\textit{lambda T.P.C., S.S., M.C.: T.P.C.* S.S. > M.C., M.C.: [$2^0, 2^1, ..., 2^9, 2^{10}$]}.
One solution to satisfy the constraint is \{T.P.C.:4, S.S.:32, M.C.:256\}.
As such, \name{} assigns 256 to \texttt{MaxClients}.
Apache is now in the state S4: \{ON,256,32,4\}. S4 has a performance measurement of 30 r/s.
The immediate reward for taking action A7 in S3 is 0.2.

In the fourth interaction, the agent receives the action A6, 
which is to decrease the value of \texttt{StartServers} (S.S.).
The \texttt{StartServers} option gets a new option value of 16.
The subject is now in the state S5: \{ON,256,16,4\}.
S5 has a performance measurement of 20 r/s. 
The immediate reward for taking action A6 in S4 is -0.17.
After the first iteration finishes,
dynamic state reduction looks through states that have identical performance 
and combines such states.

\revision{For instance, states S2 and S5 are to be combined, 
and the performance-state cache gets a new entry:
\{20: \{MasterState: S2 -> SlaveStates: S5\}.}
 
\section{Empirical Study}
\label{sec:study}
To evaluate \name{}, we conduct an empirical study on four subjects 
and aim to answer the following research questions:

\vspace*{3pt}
\noindent
\textbf{RQ1:} How effective is \name{} in
tuning the configuration options values to
achieve long-term performance gains?

\vspace*{3pt}
\noindent
\textbf{RQ2:} How efficient is \name{} 
for achieving a given performance goal?

\subsection{Implementation}
\revision{For reinforcement learning, we extend the Python-based 
library pybrain~\cite{schaul2010pybrain} to conduct Q-Learning.
The Bash shell script is used to connect and execute different components of experiments 
such as updating configurations, monitoring performance measurements, and calculating rewards.
We conduct all the experiments automatically 
on a RedHat Linux system in a High-Performance Computer (HPC) cluster. 
The basic HPC node is equipped with a 6 core 2.66 GHz Intel Xeon X5650 Westmere, 
36 GB of memory, and 256 GB of hard drive.}

\begin{table}[h]
\centering
\setlength{\tabcolsep}{4pt}
\caption{\label{tab:subs} Characteristics of Subjects}
\begin{tabular}{|c|c|c|c|} \hline
{\em Subject} & {\em Modules}
& \#{\em $O_{n}$} &\#{\em $O_{b}$} \\
\hline 
{Apache} & CORE, WORKER, MPM\_COMMON 
& {33} & {17} \\
\hline
{Lighttpd} & CORE & {10} & {23} \\  
\hline
{MySQL} & INNODB, SYSTEM 
& {29} & {24} \\ 
\hline
{PSQL} & CORE & {40} & {38} \\
\hline
\end{tabular}
\begin{center}
\end{center}
\normalsize
\end{table}

\subsection{Subject Programs}
We choose four popular open-source
server applications: Apache HTTPD Server, Lighttpd Web Server,
MySQL Server, and PostgreSQL (PSQL) Server. 
All subjects are highly configurable server applications, 
which are prone to performance issues caused by misconfigurations. 
Table~\ref{tab:subs} shows the characteristics of the subjects. 
The ``Modules'' column shows the modules from which the configuration options 
are collected.  We evaluated the modules involving the core functionalities  of the programs. 
The ``\#$O_n$'' column lists the number of numeric options 
and the ``\#$O_b$'' column lists the number of binary options.

\revision{To evaluate the performance of subject programs, 
we use the program throughput as the performance measurement 
similar to other work~\cite{bu2009reinforcement}.
Specifically, we use the number of concurrent web page requests (CWR)
for web servers and
the number of transactions per second (TPS) for database servers.
CWR and TPS are commonly used performance measurements for 
web and database servers~\cite{dbt2}.}

\subsection{Experiment Design}

\subsubsection{Baseline Techniques}

We use a random method M$_{RND}$ as the baseline for both effectiveness and efficiency 
comparison since there are no existing techniques that target the same goal as \name{}. 
M$_{RND}$ randomly selects a configuration option from the configuration 
option pool~\footnote{The configuration option pool includes all options available 
in the selected modules to the subject programs.} 
and then assigns a random value to the configuration option according to its value range.
M$_{RND}$ skips state reduction and reinforcement learning steps.
To evaluate whether the adaptive value generation and dynamic state merging techniques
can affect the effectiveness and efficiency of \name{},
we consider two ``vanilla'' versions of \name{}. 
The first version is \name{}$_A$, which does not apply dynamic state merging.
The second version is \name{}$_D$, which does not apply the adaptive value generation.
Similar to M$_{RND}$, \name{}$_D$ assigns a random value to the configuration option.  
We let each technique run for 24 hours for as many iterations as it can complete 
before checking the results.
To reduce the influence of randomness, we repeat each method 10 times.
The null-hypothesis, ${H_0}$ states that ``the mean of the two methods are equal'', 
and we reject the null hypothesis if the probability value is less than 5\% (p < 0.05).
After all methods finish, we conduct the t-test to evaluate if the mean difference 
in each set of data is statistically significant.

\subsubsection{RQ1: Effectiveness of \name{}}

RQ1 evaluates whether \name{}  
is effective at guiding the applications toward higher performance by adjusting
the configuration option values. 
Since the first step of \name{} is to identify performance-influential configuration
options to reduce the search space, we want to evaluate if 
the ranking is accurate compared to M$_{RND}$. 
\revision{Specifically, \name{} uses a 3-way covering array to conduct configuration sampling
for K-Means clustering.
The top-10 configuration performance-influential options are returned, 
such configuration options get a larger weight in reinforcement learning.
In other words, \name{} is biased towards selecting such options for performance tuning. 
M$_{RND}$, on the other hand, randomly selects 
10 configuration options.}

To measure the effectiveness of ranking, 
the mean average precision (MAP) score is used. 
\revision{MAP is a single-figure measurement of ranked retrieval results independent of the 
size of the top list~\cite{schutze2008introduction}.} 
It is designed for general ranked retrieval problems, where a query can have multiple relevant documents. 
To compute MAP, it first calculates the average precision (AP) for individual 
query $Q_i$, and then calculates the  mean of APs on 
the set of queries Q:

\begin{center}
 $\mathit{MAP = \frac{1}{|Q|} \cdot \displaystyle\sum_{Q_i \in Q} AP(Q_i)}$.
\end{center}

\revision{To illustrate the MAP calculation, suppose there are two configuration options $O_1$ and $O_2$
that are performance influential.}
If Technique-I ranks the two options 
at the 1st and 2nd positions and Technique-II ranks 
the two options at the 1st and 3rd positions, then the MAP score
of Technique-I is (1/1 + 2/2)/2 = 1 and the MAP score of Technique-II
is (1/1 + 2/3)/2 = 0.8.
\revision{We say that Technique-I is better than Technique-II in ranking 
performance-influential configuration options.}

Next, we evaluate if \name{}'s performance tuning algorithm is effective. 
\revision{To build a Q-Table, \name{} needs to obtain performance measurements. 
Benchmark tools are used to generate workload and report 
performance measurements. 
For instance, the Apache Benchmark (ab) is used to measure Apache's performance
by the following command: ``ab -n 1000 -c 10 http:localhost''.}
-n specifies the number of requests and -c specifies the level of concurrency.
The benchmark tools provide an elegant solution to generate synthetic traffic on demand.

\revision{In the experiment environment, all non-system processes are 
terminated to dedicate the system resource to the subject program and 
to reduce any other activities that may disturb the experiment. 
We configure each subject program according to the performance tuning 
guidance~\cite{ApacheTuning1,lightdPerfTune1,pgsqlPerfTune1,mysqlkey}
and benchmark 1000 times for each subject to record the subject's performance.
The best performance is selected and used as the performance goal P$_{GOAL}$.

This performance is established as the maximum performance achievable 
in the experiment environment.

We look at the mean performance achieved when each method reaches 
the time limit.}

\subsubsection{RQ2: Efficiency of \name{}}
RQ2 evaluates how long it takes for \name{} 
to achieve a given performance goal. 
We compare \name{} with three baseline techniques
(i.e., M$_{RND}$, \name{}$_A$, and \name{}$_D$) to evaluate the 
overall efficiency of \name{}.
\revision{The mean performance measurement is checked hourly.
Ideally, the mean performance measurement should be equal to P$_{GOAL}$, 
due to the internal implementation of the benchmarking tool, 
the uncertainties on the experiment environment, and the nature of 
reinforcement learning method, it is not always possible to achieve a mean
performance measurement equal to P$_{GOAL}$.
We consider \name{} converges when the mean performance measurement 
is greater than 90\% of the P$_{GOAL}$
and maintains the same level of performance to the end of the experiment.} 
Although measuring performance is the most time-consuming operation in \name{},
each method uses the same method (benchmark tool). Hence,
the time spent on such steps is comparable.  

%
\begin{table*}[h]
\centering
\caption{\label{tab:RQ1} Effectiveness and Efficiency of \name{}}
\begin{tabular}{|c|c|c|r|r|r|r|r|r|r|r|r|}
\hline
\multirow{2}{*}{\em Subject}
& \multicolumn{2}{c|}{{\em MAP}} 
& \multicolumn{4}{c|}{\em Effectiveness} 
& \multicolumn{4}{c|}{\em Efficiency} \\

& \name{} & {$M_{RND} $} &  \name{} & $M_{RND}$ & \name{}$_A$ & \name{}$_D$ 
& \name{} & $M_{RND}$ & \name{}$_A$ & \name{}$_D$\\
\hline 
{Apache} & \bf{0.59} & {0.17}
& \bf{4607 r/s} & {3540 r/s} & 4374 r/s & 4422 r/s  
& \bf{20 h} & 24+ h & 24+ h & 24+ h\\ 
\hline 

{Lighttpd} & \bf{0.7} & 0.17 
& \bf{3864 r/s} & {3094 r/s} & {3602 r/s} & {3862 r/s}   
& {21 h} & 24+ h & 24+ h & \bf{18 h}\\
\hline 

{MySQL} & {0.42} & \bf{0.64} 
& \bf{324 t/s} & {257 t/s} & {317 t/s} & {315 t/s}   
& \bf{20 h} & 24+ h & 24+ h & 24+ h\\
\hline 
{PSQL} & \bf{0.36} & {0.11}
& \bf{248 t/s} & {217 t/s} & {232 t/s} & {235 t/s}  
& \bf{21 h} & 24+ h & 24+ h & 24+ h\\
\hline 
\end{tabular}
\begin{center}
MAP: Mean Average Precision; 
r/s: requests per second; t/s: transaction per second; h: hour;
\end{center}
\normalsize
\end{table*}

\ignore{
\name{} is compared with two different versions of \name{}, 
${M_{DSR}}$ and ${M_{AVG}}$.
${M_{DSR}}$ supports the dynamic state reduction without dynamic option value adjustment.
As such, the next option value selected is based on a random value.
${M_{AVG}}$ uses option value adjustment without the dynamic state reduction support.  
The efficiency is measured by reporting the number of states observed,
the average performance reading, and the wall-clock time used
when the reinforcement learning process completes in each method. 
For \name{}, the number of duplicated states (states with same performance reading) 
identified throughout each learning iteration was reported.
}
 
\section{Results and Analysis}
\label{sec:results}

\begin{figure*}[h]
\begin{subfigure}{.45\textwidth}
  \centering
 \includegraphics[width=.9\linewidth,scale=0.1]{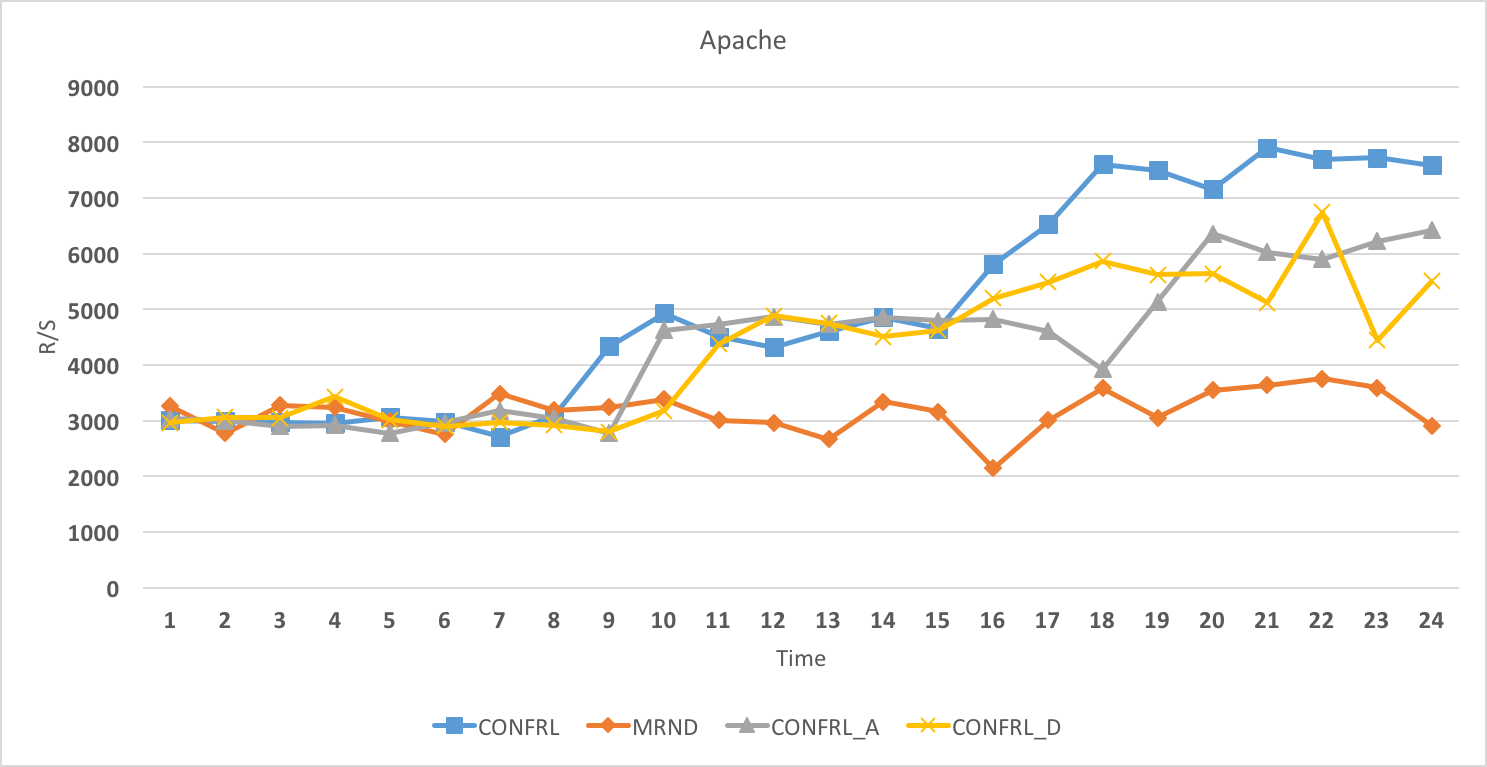}
  \caption{Apache}
  \label{fig:tpsApache}
\end{subfigure}%
\begin{subfigure}{.45\textwidth}
  \centering
   \includegraphics[width=.9\linewidth,scale=0.1]{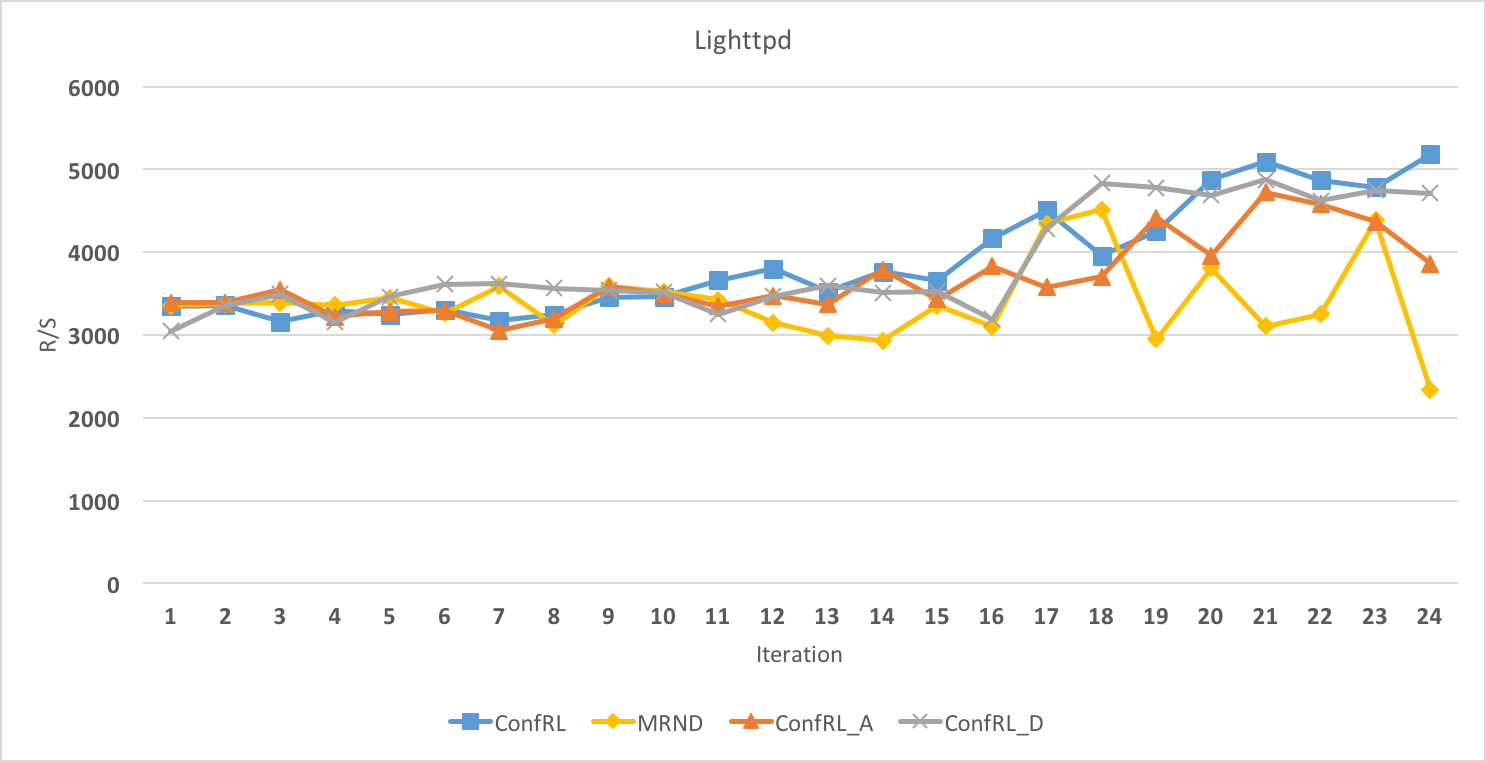}
  \caption{Lighttpd}
  \label{fig:tpsLighttpd}
\end{subfigure}

\begin{subfigure}{.45\textwidth}
  \centering
 \includegraphics[width=.9\linewidth,scale=0.1]{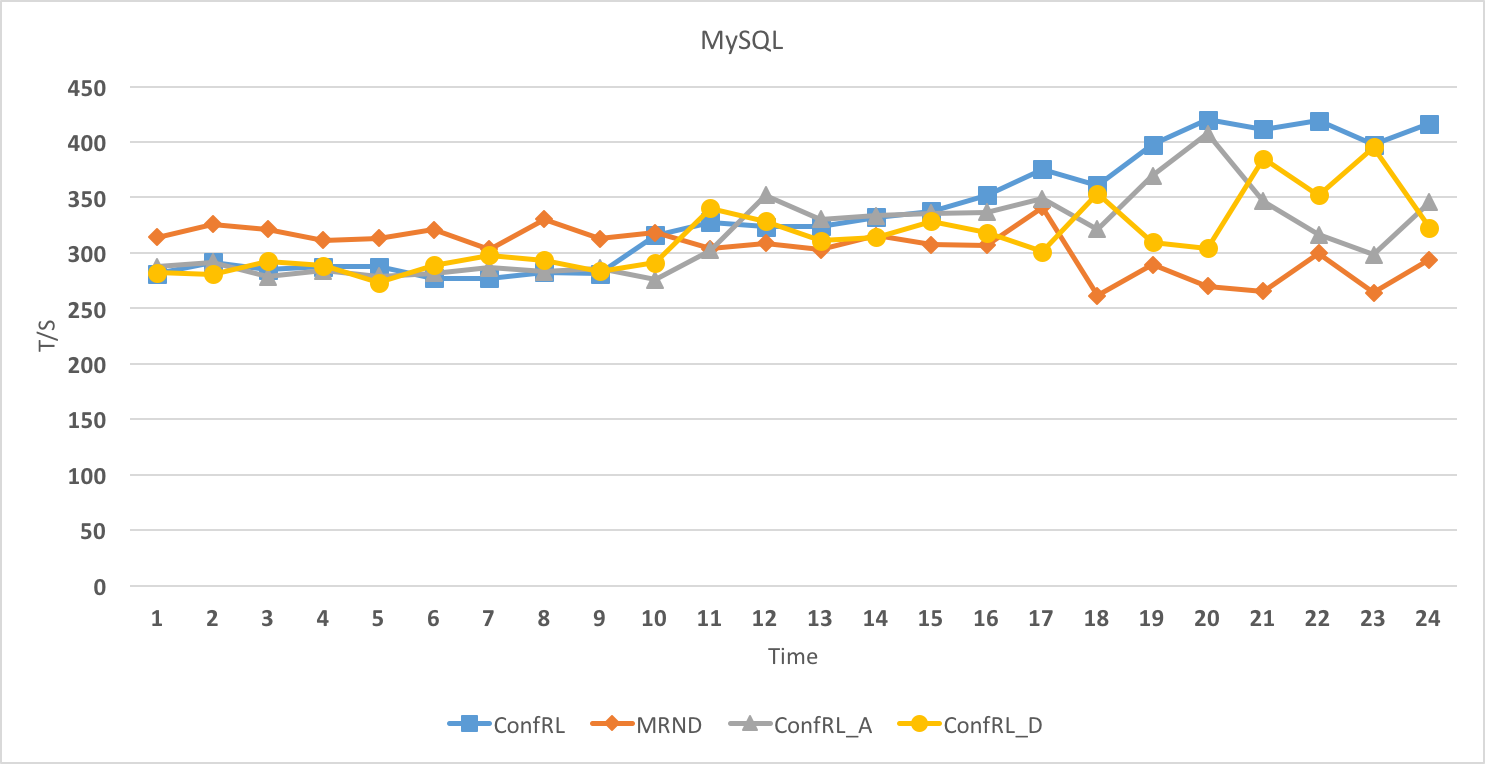}
  \caption{MySQL}
  \label{fig:tpsMysql}
\end{subfigure}%
\begin{subfigure}{.45\textwidth}
  \centering
   \includegraphics[width=.9\linewidth, scale=0.1]{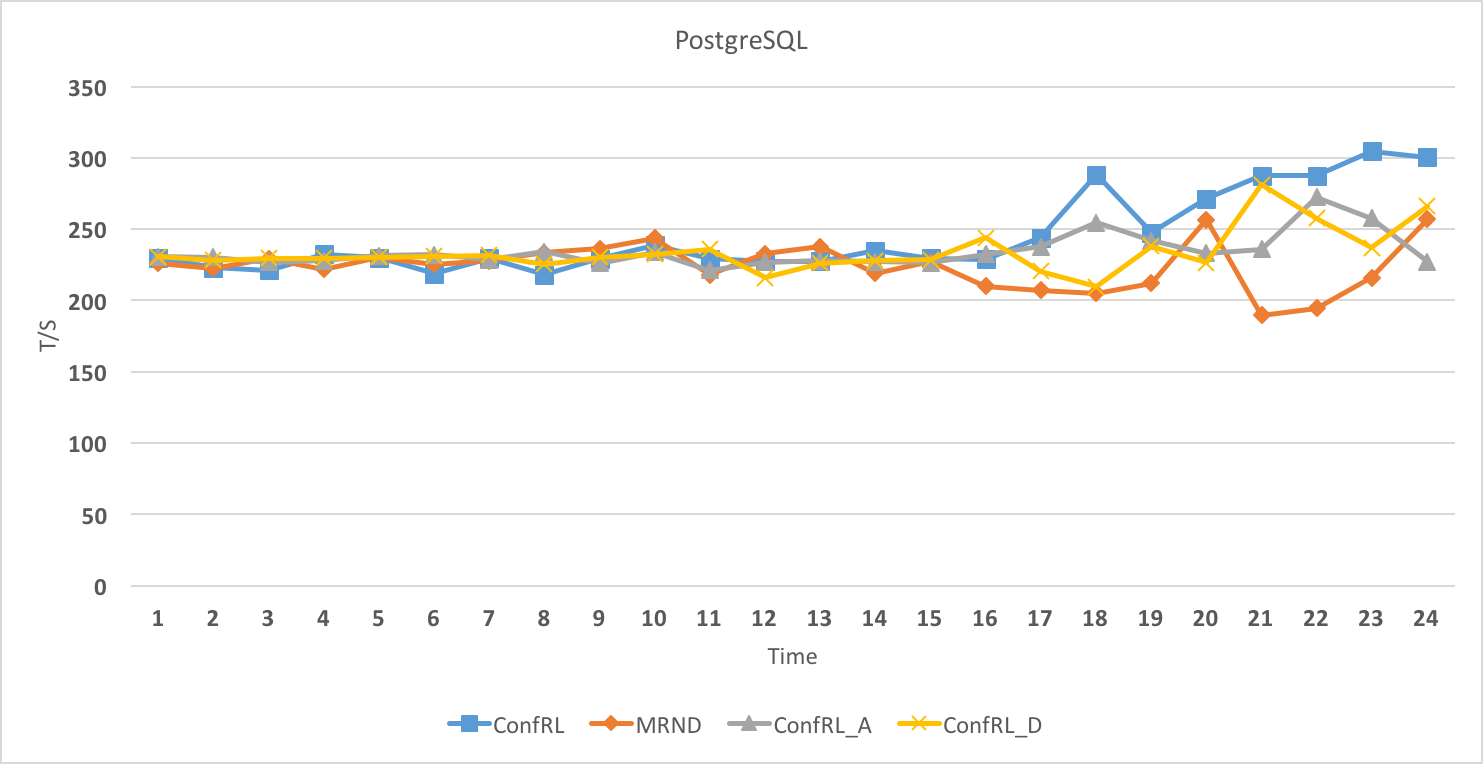}
  \caption{PostgreSQL}
  \label{fig:tpsPsql}
\end{subfigure}
\caption{Reinforcement Learning Plots}
\label{fig:plot}
\end{figure*}

\subsubsection{RQ1}

Table~\ref{tab:RQ1} shows the result of the effectiveness of ranking 
performance-influential configuration options. 
The ``MAP'' column lists the mean average precision scores for both \name{} and
M$_{RND}$.
The MAP score of \name{} ranges from 0.36 to 0.7, with
an average MAP score of 0.52. \name{} outperforms M$_{RND}$ in three 
out of four cases. 
\name{} successfully identifies at least one performance-influential option~\cite{ApacheTuning1} 
and ranks options in the top-10 position.
The results show that the option ranking method used in \name{} is effective.

In the ``Effectiveness'' column, we report
the mean performance measurements
across all the iterations for \name{}, M$_{RND}$, \name{}$_A$, and  \name{}$_D$, respectively.
As the results suggest, \name{} outperforms M$_{RND}$ in all four programs,
ranging from  14\% to 30\%, with 24\% on average. 
The t-test shows the difference between the two sets of data is statistically significant.
The result shows that \name{} can effectively select the right configuration options 
to tune performance.

Figure~\ref{fig:plot} shows the plotting of four methods in each subject program.
\revision{The plot shows how well each method works within the time limit.}
The x-axis indicates the timeline. The y-axis indicates the performance measurement.
Due to the space limitation, 
\revision{we place 24 data points in each plot to represent 
the average performance measurement calculated in each hour.
Since there is only a small subset of states that can lead to higher performance, 
initially, all four methods go hand in hand in terms of the mean performance measurement.}
As a matter of fact, M$_{RND}$ can often achieve similar and sometimes even better performance.

In early iterations, the performance fluctuations in \name{}, \name{}$_A$, and \name{}$_D$ are expected.
\revision{This is due to explorations in 
the reinforcement learning states (i.e., states represented by different combinations of 
configuration option values that as defined in Section~\ref{subsec:RLEG}). 
In the early reinforcement learning iterations, 
\name{} needs to explore as many states as possible to understand the environment. 
Specifically, for $\epsilon$-greedy exploration, 
\name{} does not follow exactly the path (i.e., policy) learned from 
previous iterations to explore more states 
and hence the performance fluctuations.}
There is a small chance that the agent strays away from the current policy.
The action is selected by following the $\epsilon$-Greedy algorithm. 
In a nutshell, $\epsilon$ determines the randomness of exploring outside of the learning comfort zone,
which allows the agent to have a chance to explore unseen states.
The agent either receives a random action in the exploration phase 
or an action by exploiting the experience.
This also prevents the agent from trapping at a local maximum. 
\name{} gradually reduces $\epsilon$ to help
the learning process converge faster.

Towards the end of the execution, the performance measurement tends to 
stabilize as the agent figures out what actions to take for a given state.
Because \name{} uses fewer states in the learning process, we observe that
\name{} converges faster than \name{}$_A$ and \name{}$_D$.
On the contrary, M$_{RND}$ does not learn from any previous interactions,
the performance of the random method does not have a noticeable improvement.
\revision{Therefore, \name{} is more effective in performance tuning than the baseline methods}

\subsubsection{RQ2}

The ``Efficiency'' column in Table~\ref{tab:RQ1} shows the efficiency of \name{}. 
When comparing \name{} to  the baseline random method M$_{RND}$, 
in three out of four subject programs, \name{} uses less time to converge to the target performance.
On average, \name{} uses 20.5 hours to converge whereas all other methods fail to converge 
within the 24-hours time limit except one occasion in Lighttpd with \name{}$_D$.

Lighttpd has the smallest number of configuration options in all four subjects.
The size of the configuration space hence the number of runtime states is smaller
compared to other subjects.
We conjecture it is the smaller size of the runtime states that leads to the \name{}$_D$
method to converge faster. 
Nonetheless, the result shows that the dynamic state reduction technique
is useful as the \name{}$_A$ method takes longer to converge.

Table~\ref{tab:RQ2B} shows the results of \name{}, \name{}$_A$, and
\name{}$_D$ for evaluating the impact of state reduction techniques. 
The result in Table~\ref{tab:RQ2B} shows the number of states 
reduced range from 11\% to 36\% and on average \name{} reduces 
reinforcement learning states by 22.8\%.
When comparing to the \name{}$_D$ method, the states reduced range from 10.7\% to 79.8\%
and on average \name{} uses 22.8\% fewer states. 
\name{} uses 26.3\% to 82.5\% (on average 62.3\%) fewer states when compared to the \name{}$_A$ method. 
As we can see, \name{} reduces the number of reinforcement learning states without losing
learning power.
\revision{The results show that the reinforcement learning 
together with state reduction techniques used in \name{} are efficient.}

%
\begin{table}[t!]
\centering
\footnotesize
\setlength{\tabcolsep}{4pt}
\caption{\label{tab:RQ2B} RQ3: Impact of State Reduction Techniques}
\begin{tabular}{|c|c|r|r|r|r|r|r|}
\hline
\multirow{2}{*}{\em Subject} 
& \multicolumn{3}{c|}{\em \name{}} 
 & \multicolumn{2}{c|} {\em \name{}$_{A}$} 
 & \multicolumn{2}{c|}{\em \name{}$_{D}$}\\
& States & MD & PM & States & PM & States & PM\\
  \hline 

Apache &  21696 & 7548 & 4607 r/s & 54846 & 4374 r/s & 45231 & 4422 r/s\\
\hline 

Lighttpd & 15370  & 1844 & 3864 r/s & 20854 & 3602 r/s & 18339 & 3862 r/s\\
\hline 

MySQL & 19045 & 2095 & 324 t/s & 99956 & 317 t/s & 76327 & 315 t/s\\
\hline 

PSQL & 23587 & 7811 & 248 t/s & 134524 & 232 t/s & 116761  & 235 t/s\\
\hline 
\end{tabular}
\begin{center}
States: \# of states identified when the RL finishes;
MD: \# of states sharing the same performance;
PM: performance measurement;\\
\end{center}
\end{table}
 
\section{Discussion}
\label{sec:discussion}

\subsection{Sensitivity of Strength}
We evaluate \name{} with different initial states. 
Specifically, we conduct an experiment with three sets of initial states: 
1) configuration options with default out of the box values; 2) configuration options with random values; 
3) configuration options with known best performance values. 
\revision{In the first two cases,  
\name{} can guide the subject programs towards better performance
after 200+ iterations.
However, in the third case in which subject programs 
are configured with known options for best performance, 
the performance learning plots show a zigzag pattern. 
\name{} first takes the subject program to a state with poor performance,
and then jumps to a state that results in a good performance.
It makes sense to the reinforcement learning algorithm, 
as this pattern would allow the agent to get more rewards in each iteration.
This shows a potential weakness in \name{} as it may not 
behave optimally when starting in the optimal state.
Experiments have shown that by giving a greater penalty to the agent may alleviate this phenomenon.}

\subsection{Threats to Validity}
\revision{The primary threat to the external validity of this work involves the representativeness 
of the selected subject programs. Other subjects may exhibit different behaviors.  
We reduce this threat to some extent by using several varieties of well studied open-source projects
from different application domains. 
For programs in the same application domain (e.g., web servers), 
we select multiple subject programs
with different implementations and varying numbers of configuration options.}

The primary threats to the internal validity of this work include possible faults in the implementation
of the proposed approach and tools that we use to perform the evaluation. 
We control this threat by testing our tools extensively and verifying their results against 
a small-scale program for which we can manually determine the correct results.
For each subject program, we start with a small set of manageable configurations to test things out
before conducting experiments on a larger scale.
The time complexity of dynamic state merging is proportional to the number of
runtime states, it could be very expensive when the subject program has an extensive number of 
runtime states. We control this threat by identifying the performance-influential configuration 
options and restricting the number of configuration option values to reduce the runtime state space.

%
\subsection{Limitations}
First, this work does not evaluate the impact of multi-layer software systems.
Instead, \name{} treats the other layers in a black-box fashion. 
\revision{For instance,
when requesting a web page from Apache, a web page could make calls
to render the dynamic content on the web page from a backend database server.
\name{} does not consider the impact of the backend database server when adjusting 
configuration options to the Apache server.
However, in our setup, we make sure other layers in a multi-layer software have the
same setup throughout the experiments.
Second, this work does not evaluate the impact on a resource sharing server where multiple programs 
can request hardware resources at the same time. The current setup assumes that the 
subject program is the only program demanding resources on the host machine.
This is a reasonable assumption since in practice many businesses would prefer to 
deploy web servers and database servers to dedicated machines.}

\section{Related Work}
\label{sec:related}

{\bf Configuration Auto Fix.}
Su et al.~\cite{su2007autobash} propose a causality dependency tracking and analysis approach
on modified Linux kernels to help users to find a solution to the configuration problem.
Swanson et al.~\cite{swanson2014beyond} design the REFRACT,
a self-adaptive framework to find workarounds to fix and prevent future configuration-induced software failures. 
Whitaker et al.~\cite{whitaker2004configuration} propose Chronus, a tool that utilizes 
user-provided probes to search through the incremental system checkpoints to find the offending states
and diagnose configuration errors that caused software functional problems. 
Unlike our approach that targets the software performance (non-functional requirement) long-term gain, 
the aforementioned methods target software failures (functional requirement) 
which are vastly different than performance issues.

\revision{
Several techniques have been proposed to find optimal configurations using learning 
techniques~\cite{diao2002optimizing,elkhodary2010fusion,hoffmann2011dynamic,liu2003online,nair2018finding}.
Diao et al.~\cite{diao2002optimizing} propose an approach to use the fuzzy controller 
to automatically tune configuration options that were known to have a concave 
upward effect to optimize response time. One big limitation with this approach is
that the method relied on the qualitative knowledge of selecting such configuration options.
Elkhodary et al.~\cite{elkhodary2010fusion} develop a general-purpose framework for
a self-adaptive system through a feature-oriented system model. 
Hoffmann et al.~\cite{hoffmann2011dynamic} use instrumentation to trace and 
adjust configuration parameters.
Liu et al.~\cite{liu2003online} conduct experiments to find the best optimization techniques to reduce
response time by adopting online optimization methods, such as Newton's Method, to configuration options in 
the Apache web server. However, hill climbing techniques based on Newton's method can be used 
to find the optimal value only when the problem has a concave upward effect on the parameter, 
therefore, limiting its adaptability.
Reinforcement learning used in \name, on the other hand, is known to solve the problem of determining 
what actions to take without requiring any prior knowledge of the environment.
Naturally, it is suitable for performance tuning.
}

{\bf Reinforcement Learning Techniques.}
Other literature~\cite{bu2009reinforcement,rao2009vconf} 
explore the use of reinforcement learning in the context of dynamically
adjusting resource allocations (e.g., CPU and Memory) on the resource sharing virtual machine environment.
Such efforts are mainly focused on optimizing the hardware-level resource configurations on the virtual machine
environment where guest systems may compete for shared resources.
In such cases, the size of the configuration space is relatively smaller 
since only a handful of resources are needed to be considered.
Also, the best practice for tuning performance on the hardware level is well established 
compared to the software level configurations.
Bu et al.~\cite{bu2009reinforcement} propose RAC, a reinforcement learning approach to 
automatically update the application configuration in response to the web traffic and 
virtual machine changes. 
%
Rao et al.~\cite{rao2009vconf} propose a reinforcement learning approach to automatically
configure resources on virtual machines (VM). In their work, the configuration space is defined 
in terms of the system resource allocations in the VM environment. 
The number of configuration options (CPU, MEM) to change is small.
The configuration space is much bigger in our subjects, for instance, 
Apache has hundreds of configuration options.
Also, the prior work focuses on managing resources on the VM-level to maximize throughput 
whereas we focus on achieving long-term performance gains on the application-level
with fixed hardware resources.

{\bf Control Theory Techniques.}
Previous literature~\cite{abdelzaher2002performance,padala2007adaptive,
wang2018understanding,zhang2005friendly} 
use the control theory to manipulate configurations. 
The control theory works particularly well when certain constraints must not be violated. 
However, the use of the control theory requires extensive knowledge of 
the underlying system and a lot of effort in the hyper-parameters tuning.
As previous performance bug empirical studies~\cite{wang2018understanding,zhang2005friendly} 
show, application-level configurations may have a great impact on the overall application performance.
Wang et al.~\cite{wang2018understanding} design SmartConf to use the control theory 
to build a prediction model for each option to maximize software performance while maintaining
the required operating constraints. 
SmartConf requires code modification whereas our method does not rely on the source code to work. 
Zhang et al.~\cite{zhang2005friendly} apply convergent control rules to design a
framework that enables friendly virtual machines that can adjust their demands based on feedback
on the hardware resource usage and availability.
Because of the differences in the project goals, authors of SmartConf agree that 
machine learning based techniques are ``better than controllers in deciding optimal settings''.
Abdelzaher et al.~\cite{abdelzaher2002performance} use a feedback control theory to achieve 
response time and throughput guarantees to different classes of clients in a general web server.
Padala et al.~\cite{padala2007adaptive} use classical control theory to 
allocate resources dynamically to meet the application-level quality of service 
in a virtual data center environment.
\revision{Our work, on the other hand, uses reinforcement learning 
to train the agent to automatically adjusting configuration option values to 
achieve long-term performance gains.}

\section{Conclusions}
\label{sec:conclusion}
\revision{Performance is crucial to the success of software systems.
While modern software provides great flexibility through
configuration options, the large number of configuration options can be 
difficult to understand and even more intimidating to setup properly.}
In this paper, we present \name{}, a reinforcement learning
approach that automatically tunes performance-influential
configuration options to achieve long-term performance gains. 
We evaluate \name{} with four large-scale server projects.
Our evaluation shows that \name{} can efficiently achieve higher long-term performance gains up to 30\%. 
Our experiment shows that \name{} can effectively reduce the number of reinforcement learning
states up to 82.5\%. On average, \name{} converges in 20.5 hours.
In the future, we plan to study additional factors that may influence 
the effectiveness and efficiency of \name{}, such as the context of the system environment.
We also plan to study if \name{} can be used to correct performance bugs caused by misconfiguration.

\clearpage
\newpage
\balance
\bibliographystyle{plain}
\begin{flushleft}
\bibliography{bib/short,bib/paper}

\begin{thebibliography}{10}

\bibitem{ab}
{Apache HTTP server benchmarking tool}, 2019.
\newblock https://httpd.apache.org/docs/2.4/programs/ab.html.

\bibitem{abdelzaher2002performance}
Tarek~F Abdelzaher, Kang~G Shin, and Nina Bhatti.
\newblock Performance guarantees for web server end-systems: A
  control-theoretical approach.
\newblock {\em IEEE transactions on parallel and distributed systems},
  13(1):80--96, 2002.

\bibitem{ApacheTuning1}
Apache performance tuning, 2017.
\newblock http://httpd.apache.org/docs/current/misc/perf-tuning.html.

\bibitem{Attariyan12}
Mona Attariyan, MIchael Chow, and Jason Flinn.
\newblock X-ray: Automating root-cause diagnosis of performance anomalies in
  production software.
\newblock In {\em OSDI}, pages 307--320, 2012.

\bibitem{bradtke1995reinforcement}
Steven~J Bradtke and Michael~O Duff.
\newblock Reinforcement learning methods for continuous-time markov decision
  problems.
\newblock In {\em Advances in neural information processing systems}, pages
  393--400, 1995.

\bibitem{bu2009reinforcement}
Xiangping Bu, Jia Rao, and Cheng-Zhong Xu.
\newblock A reinforcement learning approach to online web systems
  auto-configuration.
\newblock In {\em Distributed Computing Systems, 2009. ICDCS'09. 29th IEEE
  International Conference on}, pages 2--11. IEEE, 2009.

\bibitem{bugzilla}
{Bugzilla keyword descriptions}, 2016.
\newblock https://bugzilla.mozilla.org/describekeywords.cgi.

\bibitem{chung2006case}
I-H Chung and Jeffrey~K Hollingsworth.
\newblock A case study using automatic performance tuning for large-scale
  scientific programs.
\newblock In {\em 2006 15th IEEE International Conference on High Performance
  Distributed Computing}, pages 45--56. IEEE, 2006.

\bibitem{czerwonka2006pairwise}
J~Czerwonka, D~Butt, and C~Gens.
\newblock Pairwise testing in real word: practical extensions to test case
  generators.
\newblock In {\em Proc. of the 24th pacific northwest software quality conf},
  volume 2006, 2006.

\bibitem{dbt2}
{Database Test Suite}, 2019.
\newblock http://osdldbt.sourceforge.net/.

\bibitem{diao2002optimizing}
Yixin Diao, Joseph~L Hellerstein, and Sujay Parekh.
\newblock Optimizing quality of service using fuzzy control.
\newblock In {\em International Workshop on Distributed Systems: Operations and
  Management}, pages 42--53. Springer, 2002.

\bibitem{elkhodary2010fusion}
Ahmed Elkhodary, Naeem Esfahani, and Sam Malek.
\newblock Fusion: a framework for engineering self-tuning self-adaptive
  software systems.
\newblock In {\em Proceedings of the eighteenth ACM SIGSOFT international
  symposium on Foundations of software engineering}, pages 7--16. ACM, 2010.

\bibitem{Han16}
Xue Han and Tingting Yu.
\newblock An empirical study on performance bugs for highly configurable
  software systems.
\newblock In {\em Proceedings of the International Symposium on Empirical
  Software Engineering and Measurement}, pages 215--224, 2016.

\bibitem{hoffmann2011dynamic}
Henry Hoffmann, Stelios Sidiroglou, Michael Carbin, Sasa Misailovic, Anant
  Agarwal, and Martin Rinard.
\newblock Dynamic knobs for responsive power-aware computing.
\newblock In {\em ACM SIGARCH Computer Architecture News}, volume~39, pages
  199--212. ACM, 2011.

\bibitem{jamshidi2018learning}
Pooyan Jamshidi, Miguel Velez, Christian K{\"a}stner, and Norbert Siegmund.
\newblock Learning to sample: Exploiting similarities across environments to
  learn performance models for configurable systems.
\newblock In {\em Proceedings of the 2018 26th ACM Joint Meeting on European
  Software Engineering Conference and Symposium on the Foundations of Software
  Engineering}, pages 71--82. ACM, 2018.

\bibitem{jin2014preffinder}
Dongpu Jin, Myra~B Cohen, Xiao Qu, and Brian Robinson.
\newblock Preffinder: getting the right preference in configurable software
  systems.
\newblock In {\em Proceedings of the 29th ACM/IEEE international conference on
  Automated software engineering}, pages 151--162. ACM, 2014.

\bibitem{Jin12}
Guoliang Jin, Linhai Song, Xiaoming Shi, Joel Scherpelz, and Shan Lu.
\newblock Understanding and detecting real-world performance bugs.
\newblock In {\em Proceedings of the ACM SIGPLAN Conference on Programming
  Language Design and Implementation}, pages 77--88, 2012.

\bibitem{Jovic11}
Milan Jovic, Andrea Adamoli, and Matthias Hauswirth.
\newblock Catch me if you can: Performance bug detection in the wild.
\newblock In {\em Proceedings of the ACM SIGPLAN International Conference on
  Object Oriented Programming Systems Languages and Applications}, pages
  155--170, 2011.

\bibitem{kmeans}
k-means clustering, 2005.
\newblock https://en.wikipedia.org/wiki/K-means-clustering.

\bibitem{lightdPerfTune1}
Lighttpd performance improvements, 2016.
\newblock https://redmine.lighttpd.net/projects/1/wiki/docs-performance.

\bibitem{liu2003online}
Xue Liu, Lui Sha, Yixin Diao, Steven Froehlich, Joseph~L Hellerstein, and Sujay
  Parekh.
\newblock Online response time optimization of apache web server.
\newblock In {\em International Workshop on Quality of Service}, pages
  461--478. Springer, 2003.

\bibitem{nair2018finding}
Vivek Nair, Zhe Yu, Tim Menzies, Norbert Siegmund, and Sven Apel.
\newblock Finding faster configurations using flash.
\newblock {\em IEEE Transactions on Software Engineering}, 2018.

\bibitem{niemeyer2017python}
G~Niemeyer.
\newblock Python-constraint: Solving constraint satisfaction problems in
  python, 2017.

\bibitem{NistorMSR}
Adrian Nistor, Tian Jiang, and Lin Tan.
\newblock Discovering, reporting, and fixing performance bugs.
\newblock In {\em Proceedings of the International Conference on Mining
  Software Repositories}, pages 237--246, 2013.

\bibitem{padala2007adaptive}
Pradeep Padala, Kang~G Shin, Xiaoyun Zhu, Mustafa Uysal, Zhikui Wang, Sharad
  Singhal, Arif Merchant, and Kenneth Salem.
\newblock Adaptive control of virtualized resources in utility computing
  environments.
\newblock In {\em ACM SIGOPS Operating Systems Review}, volume~41, pages
  289--302. ACM, 2007.

\bibitem{pgsqlPerfTune1}
Tuning your postgresql server, 2015.
\newblock https://wiki.postgresql.org/wiki/Tuning-Your-PostgreSQL-Server.

\bibitem{puterman2014markov}
Martin~L Puterman.
\newblock {\em Markov decision processes: discrete stochastic dynamic
  programming}.
\newblock John Wiley \& Sons, 2014.

\bibitem{rao2009vconf}
Jia Rao, Xiangping Bu, Cheng-Zhong Xu, Leyi Wang, and George Yin.
\newblock Vconf: a reinforcement learning approach to virtual machines
  auto-configuration.
\newblock In {\em Proceedings of the 6th international conference on Autonomic
  computing}, pages 137--146. ACM, 2009.

\bibitem{dsilver15}
{UCL Course on RL}, 2015.
\newblock http://www0.cs.ucl.ac.uk/staff/d.silver/web/Teaching.html.

\bibitem{rlearning}
Reinforcement learning, 2002.
\newblock https://en.wikipedia.org/wiki/Reinforcement-learning.

\bibitem{schaul2010pybrain}
Tom Schaul, Justin Bayer, Daan Wierstra, Yi~Sun, Martin Felder, Frank Sehnke,
  Thomas R{\'C}$1/4$ckstie{\'C}, and J{\'C}$1/4$rgen Schmidhuber.
\newblock Pybrain.
\newblock {\em Journal of Machine Learning Research}, 11(Feb):743--746, 2010.

\bibitem{schutze2008introduction}
Hinrich Sch{\"u}tze, Christopher~D Manning, and Prabhakar Raghavan.
\newblock {\em Introduction to information retrieval}, volume~39.
\newblock Cambridge University Press, 2008.

\bibitem{mysqlkey}
{17 Key MySQL Config File Settings}, 2015.
\newblock
  http://www.speedemy.com/17-key-mysql-config-file-settings-mysql-5-7-proof/.

\bibitem{su2007autobash}
Ya-Yunn Su, Mona Attariyan, and Jason Flinn.
\newblock Autobash: improving configuration management with operating system
  causality analysis.
\newblock {\em ACM SIGOPS Operating Systems Review}, 41(6):237--250, 2007.

\bibitem{sutton1998introduction}
Richard~S Sutton, Andrew~G Barto, et~al.
\newblock {\em Introduction to reinforcement learning}, volume 135.
\newblock MIT press Cambridge, 1998.

\bibitem{Swanson14}
Jacob Swanson, Myra~B. Cohen, Matthew~B. Dwyer, Brady~J. Garvin, and Justin
  Firestone.
\newblock Beyond the rainbow: Self-adaptive failure avoidance in configurable
  systems.
\newblock In {\em FSE}, pages 377--388, 2014.

\bibitem{swanson2014beyond}
Jacob Swanson, Myra~B Cohen, Matthew~B Dwyer, Brady~J Garvin, and Justin
  Firestone.
\newblock Beyond the rainbow: self-adaptive failure avoidance in configurable
  systems.
\newblock In {\em Proceedings of the 22nd ACM SIGSOFT International Symposium
  on Foundations of Software Engineering}, pages 377--388. ACM, 2014.

\bibitem{valmari1996state}
Antti Valmari.
\newblock The state explosion problem.
\newblock In {\em Advanced Course on Petri Nets}, pages 429--528. Springer,
  1996.

\bibitem{wang2018understanding}
Shu Wang, Chi Li, Henry Hoffmann, Shan Lu, William Sentosa, and Achmad~Imam
  Kistijantoro.
\newblock Understanding and auto-adjusting performance-sensitive
  configurations.
\newblock In {\em Proceedings of the Twenty-Third International Conference on
  Architectural Support for Programming Languages and Operating Systems}, pages
  154--168. ACM, 2018.

\bibitem{whitaker2004configuration}
Andrew Whitaker, Richard~S Cox, Steven~D Gribble, et~al.
\newblock Configuration debugging as search: Finding the needle in the
  haystack.
\newblock In {\em Osdi}, volume~4, pages 6--6, 2004.

\bibitem{wikiQlearning}
{Q-learning}, 2004.
\newblock https://en.wikipedia.org/wiki/Q-learning.

\bibitem{xiao2013context}
Xusheng Xiao, Shi Han, Dongmei Zhang, and Tao Xie.
\newblock Context-sensitive delta inference for identifying workload-dependent
  performance bottlenecks.
\newblock In {\em Proceedings of the International Symposium on Software
  Testing and Analysis}, pages 90--100, 2013.

\bibitem{yilmaz2006covering}
Cemal Yilmaz, Myra~B Cohen, and Adam~A Porter.
\newblock Covering arrays for efficient fault characterization in complex
  configuration spaces.
\newblock {\em IEEE Transactions on Software Engineering}, 32(1):20--34, 2006.

\bibitem{zhang2005friendly}
Yuting Zhang, Azer Bestavros, Mina Guirguis, Ibrahim Matta, and Richard West.
\newblock Friendly virtual machines: leveraging a feedback-control model for
  application adaptation.
\newblock In {\em Proceedings of the 1st ACM/USENIX international conference on
  Virtual execution environments}, pages 2--12. ACM, 2005.

\end{thebibliography}
\end{flushleft}

\end{document}